\documentclass[pre,aps,twocolumn,superscriptaddress]{revtex4-2}

\usepackage{graphicx}
\usepackage{amssymb,amsfonts,amsmath}
\usepackage{color}
\usepackage{ulem}
\usepackage[hidelinks]{hyperref}
\usepackage{bbm}
\usepackage{bbold}

\usepackage{tikz}
\usetikzlibrary{shapes}
\usetikzlibrary{patterns}
\usetikzlibrary{angles, quotes}

\newcommand{\rv}{{\mathbf r}}

\newcommand{\Tr}{{\rm Tr}\,}
\newcommand{\e}{{\rm e}}

\newcommand{\pv}{{\bf p}}

\newcommand{\Fv}{{\bf F}}

\newcommand{\msphantom}[1]{$\ldots$}

\newcommand{\eps}{{\boldsymbol \epsilon}}

\newcommand{\eqr}[1]{Eq.~\eqref{#1}}

\newcommand{\unity}{{\mathbbm 1}}

\newcommand{\mydelete}[1]{{}}
\newcommand{\taub}{{\boldsymbol\tau}}

\newcommand{\rmint}{{\rm int}}

\newcommand{\rmext}{{\rm ext}}

\newcommand{\bsig}{\boldsymbol\sigma}
\newcommand{\Sv}{{\bf S}}

\begin{document}

\title{Gauge Invariance of Equilibrium Statistical Mechanics}

\author{Johanna M\"uller}
\affiliation{Theoretische Physik II, Physikalisches Institut, 
  Universit{\"a}t Bayreuth, D-95447 Bayreuth, Germany}

\author{Sophie Hermann}
\affiliation{Theoretische Physik II, Physikalisches Institut, 
  Universit{\"a}t Bayreuth, D-95447 Bayreuth, Germany}

\author{Florian Samm\"uller}
\affiliation{Theoretische Physik II, Physikalisches Institut, 
  Universit{\"a}t Bayreuth, D-95447 Bayreuth, Germany}

\author{Matthias Schmidt}
\affiliation{Theoretische Physik II, Physikalisches Institut, 
  Universit{\"a}t Bayreuth, D-95447 Bayreuth, Germany}
\email{Matthias.Schmidt@uni-bayreuth.de}

\date{27 June 2024, revised version: 16 August 2024}

\begin{abstract}
  We identify a recently proposed shifting operation on classical
  phase space as a gauge transformation for statistical mechanical
  microstates. The infinitesimal generators of the continuous gauge
  group form a non-commutative Lie algebra, which induces exact sum
  rules when thermally averaged.  Gauge invariance with respect to
  finite shifting is demonstrated via Monte Carlo simulation in the
  transformed phase space which generates identical equilibrium
  averages. Our results point towards a deeper basis of statistical
  mechanics than previously known and they offer avenues for
  systematic construction of exact identities and of sampling
  algorithms.
\end{abstract}

\maketitle

One of the arguably most important applications of Noether's theorem
of invariant variations \cite{noether1918, byers1998, brading2002,
  read2022book} is the systematic treatment of local gauge invariances
within the fundamental physical field theories for the
electromagnetic, weak, and strong interactions.  Using the
corresponding continuous gauge groups $\rm U(1)$, $\rm SU(2)$, and
$\rm SU(3)$ as fundamental building blocks for theory construction is
one of the most successful strategies in modern physics.  The nature
of the physical mechanisms that underlie the symmetry do however not
feature explicitly in Noether's theorem.  The theorem rather
constitutes a power tool to obtain exact equations, usually in the
form of global or local conservation laws, from an underlying
continuous symmetry that needs to have been identified within (or
input into) a variational formulation of the considered physics.

The roots of statistical mechanics are older than the modern gauge
field theories. Nevertheless, Noether's theorem has been applied only
relatively recently in various different productive ways to the
physics of equilibrium and nonequilibrium many-body systems
\cite{baez2013markov, marvian2014quantum, sasa2016, sasa2019,
  revzen1970, budkov2022, bravetti2023, brandyshev2023}.  The role
that exact sum rules \cite{hansen2013, evans1979, baus1984, evans1990,
  henderson1992, triezenberg1972} play in statistical mechanics is
akin to that of conservation laws in dynamical theories, in that they
allow to constrain and rationalize the nature of the physics, without
in general determining the full solution of the problem at hand.

In a range of recent investigations Noether's theorem has been applied
to a specific shifting operation on phase space
\cite{hermann2021noether, hermann2022topicalReview,
  hermann2022variance, hermann2022quantum, sammueller2023whatIsLiquid,
  hermann2023whatIsLiquid, robitschko2024any}, where instead of the
more usual conservation laws both well-known and new statistical
mechanical sum rules were obtained systematically. Thereby Noether's
concept of invariance against continuous transformation is applied to
statistical mechanical functionals, such as the partition sum. While
similarities with global spatial translational invariance, as
generates linear momentum conservation, were discussed
\cite{hermann2021noether, hermann2022topicalReview}, neither the
physical nature nor the mathematical structure of the general phase
space shifting transformation \cite{hermann2022quantum,
  sammueller2023whatIsLiquid, hermann2023whatIsLiquid,
  robitschko2024any, tschopp2022forceDFT, sammueller2022forceDFT} have
been unravelled.

Here we identify the phase space shifting transformation
\cite{hermann2021noether, hermann2022topicalReview,
  hermann2022variance, hermann2022quantum, sammueller2023whatIsLiquid,
  hermann2023whatIsLiquid, robitschko2024any} as a local gauge
symmetry transformation that is inherent to the statistical mechanics
of particle-based systems. Realizing the defining feature of a gauge
transformation, the application of the local shifting has no effect on
any physical observables. Despite the shifting being geometric, the
transformation is non-commutative, even when displacing only
infinitesimally.  A non-commutative Lie algebra of generators
characterizes infinitesimal transformations. Corresponding exact sum
rules follow for thermal averages. Finite transformations retain the
gauge invariance as we demonstrate via Monte Carlo computer
simulations.

The shifting operation put forward in Refs.~\cite{hermann2021noether,
  hermann2022topicalReview, hermann2022quantum, hermann2022variance,
  sammueller2023whatIsLiquid, hermann2023whatIsLiquid,
  robitschko2024any} affects the positions $\rv_i$ and momenta $\pv_i$
of each particle~$i=1,\ldots, N$ via the following transformation:
\begin{align}
  \rv_i &\to \rv_i + \eps(\rv_i) = \tilde \rv_i,
  \label{LQriTrafo}\\
  \pv_i &\to [\unity + \nabla_i\eps(\rv_i)]^{-1}\cdot \pv_i = \tilde \pv_i,
  \label{LQpviTrafo}
\end{align}
where the $d$-dimensional vector field $\eps(\rv_i)$ is such that
\eqr{LQriTrafo} is a diffeomorphism, i.e., together with its inverse
is bijective and smooth; $d$ is the spatial dimensionality and the
tilde indicates the new phase space variables.  In \eqr{LQpviTrafo}
the symbol $\unity$ denotes the $d\times d$ unit matrix, $\nabla_i$ is
the derivative with respect to $\rv_i$, and the superscript $-1$
denotes matrix inversion. The transformation is canonical in the sense
of classical mechanics~\cite{goldstein2002} and hence the differential
phase space volume element is preserved, $d\rv_id\pv_i=d\tilde\rv_i
d\tilde \pv_i$. This property is fundamental for thermal averages to
arise as invariant under the application of Eqs.~\eqref{LQriTrafo} and
\eqref{LQpviTrafo}.

To be specific, we consider the statistical mechanics of
Hamiltonians~$H$ with the standard form
\begin{align}
  H &= \sum_i\frac{\pv_i^2}{2m} + u(\rv^N) + \sum_i V_\rmext(\rv_i),
  \label{LQHamiltonian}
\end{align}
where the sums run over all $N$ particle indices $i$, $m$ denotes the
particle mass, $u(\rv^N)$ is the interparticle interaction potential,
and $V_\rmext(\rv)$ is an external one-body potential.  We use the
shorthand notation $\rv^N=\rv_1,\ldots,\rv_N$ and
$\pv^N=\pv_1,\ldots,\pv_N$ to indicate the phase space variables of
all particles.  The statistical mechanics is based on the grand
ensemble with chemical potential $\mu$ and temperature~$T$. The grand
partition sum is $\Xi=\Tr\e^{-\beta(H-\mu N)}$, where the classical
trace is defined as $\Tr\cdot=\sum_{N=0}^\infty(N!h^{dN})^{-1}\int
d\rv^N d\pv^N \cdot$, with $\int d\rv^N d\pv^N$ denoting the phase
space integral over the position and momentum coordinates of all $N$
particles, $\beta=1/(k_BT)$, and $k_B$ denoting the Boltzmann
constant. The grand potential is $\Omega=-k_BT\ln\Xi$ and thermal
averages are obtained as $\langle \cdot \rangle = \Tr \cdot \e^{-\beta
  (H-\mu N)}/\Xi$.

We here introduce operator methods to capture the essence of the phase
space shifting \eqref{LQriTrafo} and \eqref{LQpviTrafo}. Specifically
we define the following, at each position $\rv$ localized, phase space
shifting operators:
\begin{align}
  \bsig(\rv) &= \sum_i \big[
    \delta(\rv-\rv_i)\nabla_i 
    +\pv_i\nabla\delta(\rv-\rv_i)\cdot\nabla_{\pv_i}
    \big],
  \label{LQsigma}
\end{align}
where $\delta(\cdot)$ denotes the Dirac distribution in $d$
dimensions, $\nabla$ indicates the derivative with respect to position
$\rv$, $\nabla_{\pv_i}$ is the momentum derivative with respect to
$\pv_i$, and we recall that $\nabla_i$ is the derivative with respect
to $\rv_i$. The shifting operators~\eqref{LQsigma} possess two key
properties. First $\boldsymbol\sigma(\rv)$ is anti-self-adjoint on
phase space:
\begin{align}
    \boldsymbol\sigma^\dagger(\rv)=-\boldsymbol\sigma(\rv).
\label{LQsigmaAntiSelfAdjoint}
\end{align}
The adjoint operator is indicated by the dagger and it has the
standard definition: $\int d\rv^N d\pv^N f \bsig(\rv) g= \int d\rv^N
d\pv^N g\bsig^\dagger(\rv) f$ for arbitrary phase space functions
$f(\rv^N, \pv^N)$ and $g(\rv^N, \pv^N)$. Equation
\eqref{LQsigmaAntiSelfAdjoint} is readily proven via phase space
integration by parts and the product rule (for $f$ and $g$ being
well-behaved).

Secondly, the consecutive action of two shifting operators that are
respectively localized at positions $\rv$ and $\rv'$ satisfies the
commutator relation:
\begin{align}
  [\bsig(\rv),\bsig(\rv')] &= 
  \bsig(\rv')[\nabla\delta(\rv-\rv')] + [\nabla\delta(\rv-\rv')]\bsig(\rv).
  \label{LQsigmaAlgebra}
\end{align}
We have used the standard definition of commutators of vectors:
$[\boldsymbol\sigma(\rv),\boldsymbol\sigma(\rv')]=
\boldsymbol\sigma(\rv) \boldsymbol\sigma(\rv')
-\boldsymbol\sigma(\rv') \boldsymbol\sigma(\rv)^{\sf T}$, where the
superscript $\sf T$ denotes matrix transposition, such that the
(Cartesian) $ab$-component is $[\sigma_a(\rv),\sigma_b(\rv')]=
\sigma_a(\rv) \sigma_b(\rv') -\sigma_b(\rv')\sigma_a(\rv)$.  Equation
\eqref{LQsigmaAlgebra} follows from explicit calculation via applying
the sequence of two shifting operators \eqref{LQsigma} and
simplifying.  It is also straightforward to show that the commutator
\eqref{LQsigmaAlgebra} is anti-self-adjoint: $[\bsig(\rv),
  \bsig(\rv')]^\dagger = -[\bsig(\rv), \bsig(\rv')]$, as is a general
property of the commutator of two anti-self-adjoint operators.
Furthermore the commutator \eqref{LQsigmaAlgebra} is anti-symmetric:
$[\bsig(\rv), \bsig(\rv')] = -[\bsig(\rv'), \bsig(\rv)]^{\sf T}$, and
it satisfies the Jacobi identity:
$[\sigma_a(\rv),[\sigma_b(\rv'),\sigma_c(\rv'')]]
+[\sigma_b(\rv'),[\sigma_c(\rv''),\sigma_a(\rv)]]
+[\sigma_c(\rv''),[\sigma_a(\rv),\sigma_b(\rv')]]=0$, as can be be
proven by explicit calculation.

The above set of distinctive properties of $\bsig(\rv)$ is closely
connected to a Lie algebra structure of infinitesimal phase space
shifting, as we lay out in the following. That the operators
\eqref{LQsigmaAlgebra} represent infinitesimal versions of the phase
space shifting according to \eqref{LQriTrafo} and \eqref{LQpviTrafo}
can be seen by multiplying with a given shifting field $\eps(\rv)$ and
integrating over $\rv$ to generate an operator $\Sigma[\eps]=\int d\rv
\eps(\rv)\cdot \bsig(\rv)$ that shifts according to the given form of
$\eps(\rv)$.  Using $\bsig(\rv)$ in the form \eqref{LQsigma} and
integrating gives
\begin{align}
  \Sigma[\eps] = \sum_i\big\{
    \eps(\rv_i)\cdot\nabla_i - 
    [\nabla_i\eps(\rv_i)]
    :\pv_i\nabla_{\pv_i}
    \big\}.
    \label{LQSigmaOperator}
\end{align}
The colon in \eqr{LQSigmaOperator} indicates a double tensor
contraction (trace of the product of the two matrices) and the phase
space shifting operator $\Sigma[\eps]$ depends functionally on the
shifting field $\eps(\rv)$ as is indicated by the brackets.

By construction a given phase space function $f(\rv^N,\pv^N)$ is
transported from $\rv^N,\pv^N$ to $\tilde\rv^N, \tilde\pv^N$ to first
order in $\eps(\rv_i)$ and $\nabla_i \eps(\rv_i)$
 by applying
\eqr{LQSigmaOperator} according to:
\begin{align}
  f(\tilde\rv^N, \tilde\pv^N) 
  &= f(\rv^N,\pv^N) + \Sigma[\eps]f(\rv^N,\pv^N).
  \label{LQfShiftedBySigma}
\end{align}
Here the phase space variables with and without tilde are related by
the transformation~\eqref{LQriTrafo} and \eqref{LQpviTrafo}.

\begin{figure}[!t]
  \vspace{1mm}
  \includegraphics[width=.99\columnwidth]{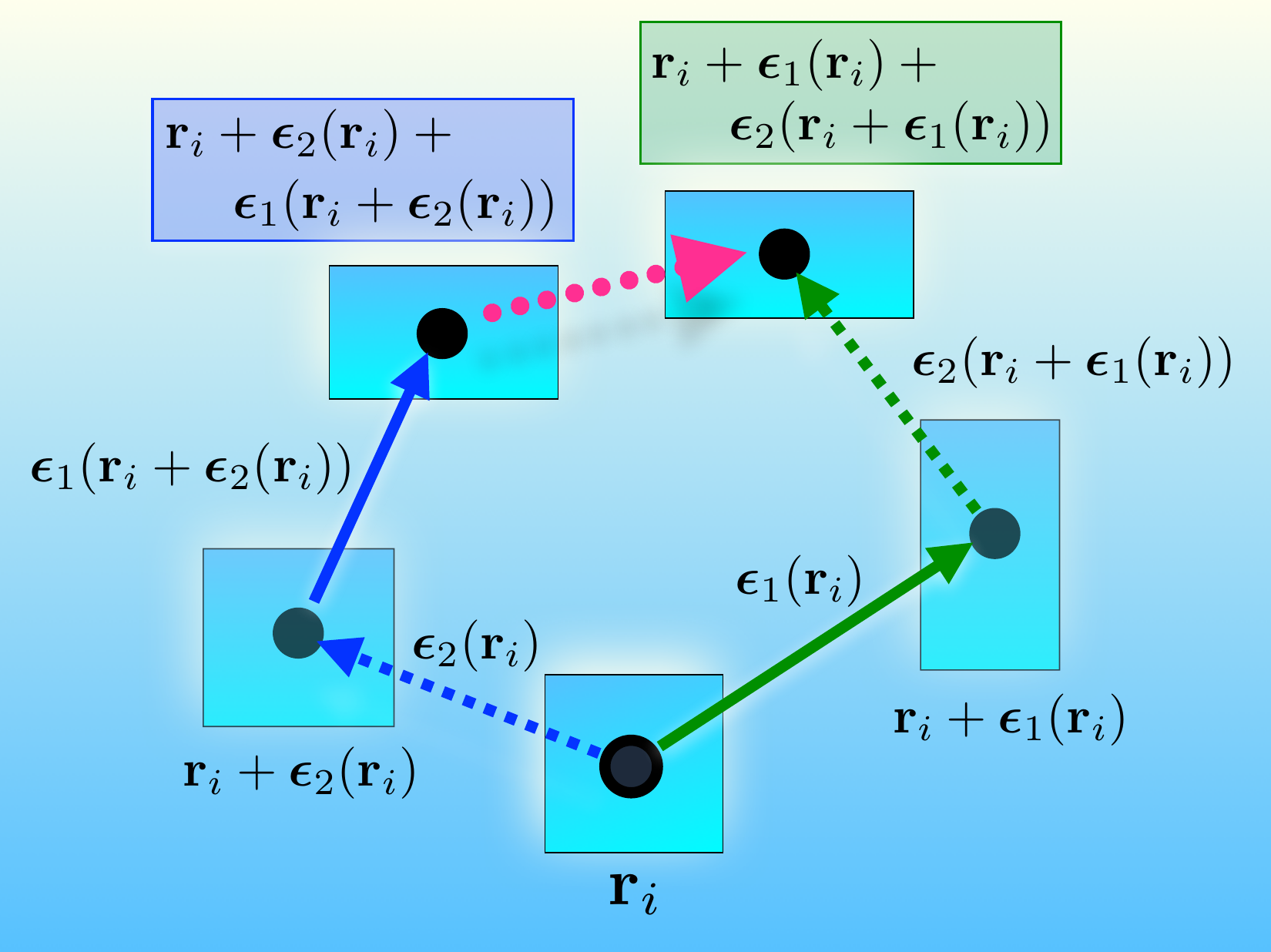}
  \caption{Illustration of the non-commutative nature of phase space
    shifting. Given are two different shifting vector fields
    $\eps_1(\rv_i)$ (solid arrows) and $\eps_2(\rv_i)$ (dashed
    arrows). Starting from $\rv_i$, the first shift yields the
    intermediate position $\rv_i+\eps_1(\rv_i)$. At this point the
    second shifting field $\eps_2(\rv_i+\eps_1(\rv_i))$ is used to
    give $\rv_i+\eps_1(\rv_i)+\eps_2(\rv_i+\eps_1(\rv_i))$ as the
    final position (two green arrows). Applying the opposite order of
    shifting and starting again at $\rv_i$ gives a different
    intermediate point $\rv_i+\eps_2(\rv_i)$.  Evaluating
    $\eps_1(\rv_i+\eps_2(\rv_i))$ at this position yields
    $\rv_i+\eps_2(\rv_i)+\eps_1(\rv_i+\eps_2(\rv_i))$ as the final
    location (two blue arrows). The two final destinations mismatch in
    general (red dotted arrow) while the differential phase space
    volume element is conserved (boxes). }
\label{FIGnoncommutative}
\end{figure}

Applying two consecutive shifts with respective vector fields
$\eps_1(\rv_i)$ and $\eps_2(\rv_i)$ constitutes repeated application
of the shifting operator according to
$\Sigma[\eps_2]\Sigma[\eps_1]$. As thereby $\Sigma[\eps_2]$ acts on
the already displaced phase space function $\Sigma[\eps_1] f(\rv^N,
\pv^N)$, the order of consecutive shifting is relevant. The
non-commutative nature of the displacements is illustrated in
Fig.~\ref{FIGnoncommutative}.

The commutator of two shifting operators $\Sigma[\eps_1]$ and
$\Sigma[\eps_2]$ quantifies the degree of their non-commutativity. Via
explicit calculation on the basis of \eqr{LQSigmaOperator} one obtains
straightforwardly the identity
\begin{align}
  [\Sigma[\eps_1], \Sigma[\eps_2]] &= \Sigma[\eps_\Delta].
  \label{LQSigmaCommutator}
\end{align}
The difference shifting vector field $\eps_\Delta(\rv_i)$ is thereby
obtained from the given forms of $\eps_1(\rv_i)$ and $\eps_2(\rv_i)$
via
\begin{align}
  \eps_\Delta(\rv_i) 
  &= \eps_1(\rv_i)\cdot[\nabla_i\eps_2(\rv_i)]
  -\eps_2(\rv_i)\cdot[\nabla_i\eps_1(\rv_i)].
  \label{LQepsilonDelta}
\end{align}
The right hand side of \eqr{LQepsilonDelta} constitutes the standard
form \cite{robbin2022book} of the Lie bracket
$[\eps_1(\rv_i),\eps_2(\rv_i)]_{\rm L}$ of the two vector fields
$\eps_1(\rv_i)$ and $\eps_2(\rv_i)$. Via replacing the functional
argument $\eps_\Delta(\rv_i)$ on the right hand side of
\eqr{LQSigmaCommutator} by the Lie bracket, we can hence alternatively
express \eqr{LQSigmaCommutator} compactly as
$[\Sigma[\eps_1],\Sigma[\eps_2]] =\Sigma[[\eps_1,\eps_2]_{\rm L}]$.

The relationship \eqref{LQSigmaCommutator} constitutes a
non-commutative Lie algebra of infinitesimal generators $\Sigma[\eps]$
due to three properties: i) anti-symmetry, ii) bilinearity, as
respectively follows from the definition of the commutator and from
linearity of the differential operator \eqref{LQSigmaOperator}, and
iii) the Jacobi identity: $[\Sigma_1, [\Sigma_2,\Sigma_3]] +
[\Sigma_2,[\Sigma_3,\Sigma_1]] + [\Sigma_3, [\Sigma_1,\Sigma_2]]=0$,
as can be verified by explicit calculation on the basis of
Eqs.~\eqref{LQSigmaOperator}--\eqref{LQepsilonDelta}; we have used the
shorthand notation $\Sigma_1=\Sigma[\eps_1]$,
$\Sigma_2=\Sigma[\eps_2]$, and $\Sigma_3=\Sigma[\eps_3]$.

The variational strategy of Refs.~\cite{hermann2022quantum,
  sammueller2023whatIsLiquid, hermann2023whatIsLiquid,
  robitschko2024any} is based on eliminating explicit occurrences of
the shifting fields. This is inline with their present status as mere
gauge functions. Within the functional calculus
methods~\cite{hermann2022quantum, sammueller2023whatIsLiquid,
  hermann2023whatIsLiquid, robitschko2024any}, one takes appropriate
functional derivatives, $\delta/\delta\eps(\rv)$, of relevant thermal
averages and then sets the shifting field to zero,
$\eps(\rv)=0$. Applying the concept to the present operator formalism
leads to differentiating the shifting operator $\Sigma[\eps]$
functionally with respect to $\eps(\rv)$. Calculating $\delta
\Sigma[\eps]/\delta\eps(\rv)= \boldsymbol\sigma(\rv)$ on the basis of
\eqr{LQSigmaOperator} is straightforward and reproduces $\bsig(\rv)$
as defined via \eqr{LQsigma}.  The functional derivative creates
spatial localization via the Dirac distribution, as it emerges from
the chain rule and the identity $\delta\eps(\rv_i)/\delta\eps(\rv)
=\delta(\rv-\rv_i)\unity$.  Due to the linearity of
\eqr{LQSigmaOperator} in $\eps(\rv)$ the dependence on $\eps(\rv)$ has
disappeared in \eqr{LQsigma}.  For how the action of $\bsig(\rv)$ is
related to differentiating by $\eps(\rv)$ see Appendix
\ref{SECrelationshipToFunctionalMethods}.

As an initial demonstration of the prowess of the localized shifting
operator \eqref{LQsigma} we apply it to the Hamiltonian with the
result
\begin{align}
  -\boldsymbol\sigma(\rv) H &= \hat \Fv(\rv),
  \label{LQFhatFromSigma}
\end{align}
where $\hat\Fv(\rv)$ is the (total) force density phase space
function, which consists of kinetic, interparticle, and external
contributions according to \cite{schmidt2022rmp}:
\begin{align}
  \hat\Fv(\rv) &= \nabla\cdot\hat\taub(\rv) + \hat \Fv_\rmint(\rv)
  - \hat\rho(\rv) \nabla V_\rmext(\rv).
  \label{EQLFviasigma}
\end{align}
The three terms on the right hand side represent kinetic stress
density $\hat\taub(\rv)=-\sum_i \pv_i\pv_i\delta(\rv-\rv_i)/m$,
interparticle force density $\hat\Fv_\rmint(\rv)=-\sum_i
\delta(\rv-\rv_i)\nabla_i u(\rv^N)$ and particle density
$\hat\rho(\rv)=\sum_i \delta(\rv-\rv_i)$ as phase space functions.

As a prerequisite for applying the localized shifting operator
approach $\bsig(\rv)$ to the thermal physics, we consider its effect
on the Boltzmann factor:
\begin{align}
  \bsig(\rv) \e^{-\beta H} &=  \beta \hat\Fv(\rv) \e^{-\beta H}.
  \label{LQsigmaToBoltzmannFactor}
\end{align}
The result \eqref{LQsigmaToBoltzmannFactor} follows from applying the
phase space derivatives in $\bsig(\rv)$ as given in \eqr{LQsigma} to
the exponential, using the chain rule, and then generating $\hat
\Fv(\rv)$ via \eqr{LQFhatFromSigma}. Applying $\bsig(\rv)$ to the
entire grand ensemble probability distribution $\e^{-\beta(H-\mu
  N)}/\Xi$ gives no additional terms, as the partition sum $\Xi$ is
not a phase space function and is hence unaffected by the action of
$\bsig(\rv)$ and the presence of the chemical potential contribution
$\e^{\beta\mu N}$ also has no adverse effect. Hence in full analogy to
\eqr{LQsigmaToBoltzmannFactor}, $\bsig(\rv) \e^{-\beta(H-\mu N)}/\Xi =
\beta \hat \Fv(\rv) \e^{-\beta (H-\mu N)}/\Xi$.

We are now ready to apply the operator algebra to the thermal physics.
We first demonstrate how prior results follow from the framework and
hence start with the thermal average $\langle \bsig(\rv) \rangle=\Tr
\bsig(\rv) \e^{-\beta(H-\mu N)}/\Xi=\langle \beta \hat \Fv(\rv)
\rangle$, as is readily obtained from \eqr{LQsigmaToBoltzmannFactor}.
On the other hand the anti-self-adjoint property
\eqref{LQsigmaAntiSelfAdjoint} allows, upon inserting a factor 1
before the shifting operator, to conclude $\langle \bsig(\rv) \rangle
= \langle 1\bsig(\rv) \rangle = \langle [\bsig^\dagger(\rv) 1] \rangle
= -\langle [\bsig(\rv)1] \rangle= -\langle 0 \rangle= 0$. Hence
overall $\langle \hat \Fv(\rv)\rangle=0$, which is the exact
equilibrium one-body force balance \cite{yvon1935, born1946,
  hansen2013, schmidt2022rmp, hermann2022quantum, tschopp2022forceDFT,
  sammueller2022forceDFT}.

Higher-order identities follow with similar ease. Consider the
two-point case, where $\langle \bsig(\rv')\bsig(\rv) \rangle=0$, which
follows as above from the adjoint $\bsig^\dagger(\rv') 1=0$.
Consecutively applying two operators yields in a first step $\langle
\bsig(\rv') \bsig(\rv) \rangle= \langle \bsig(\rv') \beta \hat
\Fv(\rv) \rangle$ according to \eqr{LQsigmaToBoltzmannFactor}. In the
second step applying $\bsig(\rv')$, using the product rule, and
bearing in mind that the overall result vanishes, one obtains the sum
rule $\beta \langle \Fv(\rv') \Fv(\rv) \rangle = -\langle [\bsig(\rv')
  \hat \Fv(\rv)] \rangle$, as previously identified in
Refs.~\cite{sammueller2023whatIsLiquid, hermann2023whatIsLiquid}. The
term on the right hand side is the mean negative force gradient or
equivalently the mean Hessian of the Hamiltonian, $-\langle
[\bsig(\rv') \hat \Fv(\rv)] \rangle = \langle [\bsig(\rv')\bsig(\rv)
  H] \rangle = \langle \delta^2 H(\tilde \rv^N, \tilde
\pv^N)/[\delta\eps(\rv)\delta\eps(\rv')]|_{\eps=0} \rangle$, as
re-written first via \eqr{LQFhatFromSigma} and then via the functional
derivative identity \eqr{LQdoublesigmaAsFunctionalDerivative} given in
appendix \ref{SECrelationshipToFunctionalMethods} and using that
$\langle \hat \Fv(\rv)\rangle=0$.

The involved force-force and force-gradient correlation functions were
shown to be highly useful measures for the spatial two-body structure
of a wide variety of soft matter systems
\cite{sammueller2023whatIsLiquid, hermann2023whatIsLiquid}. 
Similarly, by introducing a further physical observable $\hat
A(\rv^N,\pv^N)$ of interest \cite{hirschfelder1960,
  robitschko2024any}, we consider $\langle \bsig(\rv) \hat
A\rangle=0$, which follows again from using the
adjoint~\eqref{LQsigmaAntiSelfAdjoint}. Applying the shifting operator
to the functions on its right gives $\langle \beta \hat\Fv(\rv) \hat A
\rangle + \langle [\bsig(\rv) \hat A] \rangle=0$, which is the recent
general hyperforce sum rule by Robitschko {\it et
  al.}~\cite{robitschko2024any}.
\begin{figure}[!t]
  \includegraphics[width=.95\columnwidth]{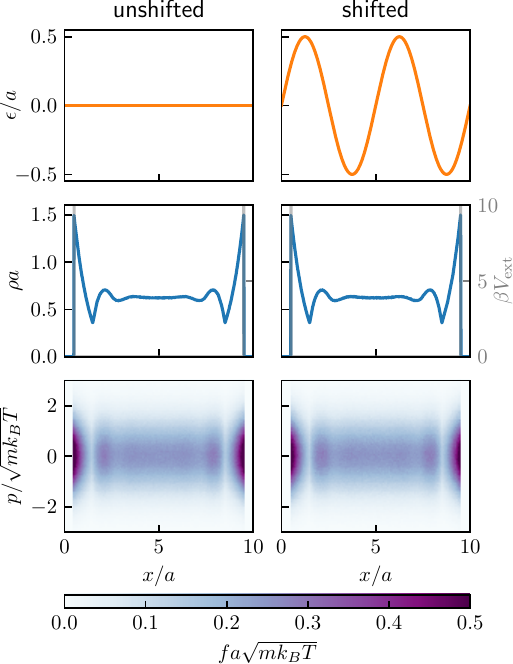}
  \caption{Monte Carlo results for one-dimensional hard rods confined
    between two hard walls with separation distance $L=10a$. Shown are
    results in the unshifted (left column) and shifted systems (right
    column). The shifting field $\epsilon$ (top panels) displaces both
    coordinates and momenta. Despite the different sampling and Markov
    chain the scaled density profile $\rho(x)a$ (middle panels) and
    one-body phase space distribution function $f(x,p)a\sqrt{mk_BT}$
    (bottom panels) remain numerically identical.
\label{FIGshiftingMC}}
\end{figure}
We re-write the second term as the thermal average $\Sv_A(\rv)=\langle
\hat \Sv_A(\rv) \rangle$ where we have defined the hyperforce phase
space function \cite{robitschko2024any} as $\hat\Sv_A(\rv)=
[\bsig(\rv) \hat A]$ (which is explicitly $\hat \Sv_A(\rv)=\sum_i
\delta(\rv-\rv_i)\nabla_i\hat A$ in case $\hat A$ is independent of
momenta).  The one-body hyperforce sum rule \cite{robitschko2024any}
can then be expressed in the compact form
\begin{align}
  \langle \beta \hat \Fv(\rv)\hat A
  \rangle + \Sv_A(\rv) &= 0. 
\label{LQhyperForceSumRuleOneBody}
\end{align}

Repeated application of the localized shifting can be realized via
similar steps as described above. Sketching a typical case for two
shifts, we have $0=\langle [\bsig^\dagger(\rv')1]\hat A
\bsig(\rv)\rangle = \langle \bsig(\rv') \hat A \bsig(\rv) \rangle =
\langle [\bsig(\rv') \hat A]\bsig(\rv)\rangle + \langle \hat A
\bsig(\rv') \bsig(\rv)\rangle = \langle [\bsig^\dagger(\rv)
  \bsig(\rv')\hat A]^{\sf T}\rangle +\langle \hat A
\bsig(\rv')\bsig(\rv) \rangle$.  The very last term can be made more
explicit as $\langle \hat A \bsig(\rv')\bsig(\rv)\rangle= \beta^2
\langle \hat A \hat \Fv(\rv') \hat \Fv(\rv) \rangle - \beta \langle
\hat A [\bsig(\rv')\bsig(\rv) H]\rangle$, which gives upon using
\eqr{LQsigmaAntiSelfAdjoint} and re-arranging the overall result
$\beta^2 \langle \hat A \hat\Fv(\rv') \hat \Fv(\rv)\rangle = \beta
\langle \hat A [\bsig(\rv') \bsig(\rv)H]\rangle + \langle [\bsig(\rv)
  \bsig(\rv')\hat A]^{\sf T} \rangle$, which is the two-body
hyperforce sum rule of Ref.~\cite{robitschko2024any} (the shifting
field in Ref.~\cite{robitschko2024any} is set to zero after each
individual functional derivative is taken).

That these exact correlation identities emerge with relatively little
effort from the present operator formalism points to its relevance.
Besides technical efficacy the formalism however allows to reveal the
rich additional structure that is generated by the Lie algebra
\eqref{LQSigmaCommutator}.  As a demonstration we multiply
\eqr{LQsigmaAlgebra} from the left by $\hat A(\rv^N,\pv^N)$ and then
build the thermal average. Writing out the resulting sandwich
structure of the integral gives for the first term on the left hand
side $\langle \hat A \bsig(\rv) \bsig(\rv') \rangle = -\langle [
  \bsig(\rv) \hat A] \beta \Fv(\rv')\rangle = -\langle \hat\Sv_A(\rv)
\beta \hat\Fv(\rv') \rangle$. Similar treatment of the second terms
yields the following exact Lie sum rule:
\begin{align}
  &    \langle \hat \Sv_A(\rv) \beta \hat \Fv(\rv') \rangle
  -\langle \beta \hat \Fv(\rv) \hat \Sv_A(\rv') \rangle
  \notag \\ & \qquad
  = \Sv_A(\rv') [\nabla\delta(\rv-\rv')]
   + [\nabla\delta(\rv-\rv')] \Sv_A(\rv).
  \label{LQsumRuleLie}
\end{align}
The right hand side of \eqr{LQsumRuleLie} follows from the average of
the product of $\hat A$ with the right hand side of
\eqr{LQsigmaAlgebra} and noting that $\langle \hat A \bsig(\rv)
\rangle=\langle [-\bsig(\rv) \hat A]\rangle=-\Sv_A(\rv)$.  The right
hand side of \eqr{LQsumRuleLie} can alternatively be re-written via
\eqr{LQhyperForceSumRuleOneBody}.

The significance of the sum rule \eqref{LQsumRuleLie} is that it
imprints the structure of the Lie operator algebra
\eqref{LQsigmaAlgebra} onto measurable spatial correlation
functions. There are two immediate consequences. First, for the case
$\rv\neq \rv'$, the right hand side of \eqr{LQsumRuleLie} vanishes and
the following nontrivial exchange symmetry emerges:
\begin{align}
  \langle \hat \Sv_A(\rv) \hat\Fv(\rv') \rangle
  &= \langle \hat \Fv(\rv) \hat \Sv_A(\rv') \rangle.
  \label{LQhyperForceExchangeSymmetry}
\end{align}
Equation \eqref{LQhyperForceExchangeSymmetry} implies the invariance
of the correlation against exchange of the force and hyperforce
densities at two distinct positions~$\rv$ and $\rv'$.

Secondly, the singular (``self'') contribution that occurs for
$\rv=\rv'$ in \eqr{LQsumRuleLie} does not generate any new one-body
correlation functions. Rather, besides the gradient of the delta
distribution, the one-body hyperforce correlation function
$\Sv_A(\rv)$, as it appears in the one-body hyperforce sum rule
\eqref{LQhyperForceSumRuleOneBody}, re-emerges. Hence the present
example demonstrates both that the Lie algebra \eqref{LQsigmaAlgebra}
i) systematically interrelates the different $n$-body levels of
correlation identities and ii) that, despite possible re-writings of
equivalent expressions, the set of relevant correlation functions that
is associated with a given observable $\hat A$ is closed.

We have thus far considered the infinitesimal structure of phase space
shifting.  In standard treatments using the exponential map one
generates a Lie group of finite transformations from a given Lie
algebra \cite{robbin2022book}. Here we use an alternative route to
demonstrate directly the invariance of the thermal physics under the
general transformation~\eqref{LQriTrafo} and \eqref{LQpviTrafo}. Via
particle-based Monte Carlo simulations we demonstrate the physical
reality of the gauge invariance by considering finite shifting which
we perform numerically. We continue to work with the full phase space
variables (as are also relevant for equilibrium Molecular Dynamics)
and hence resolve both position and momentum.

As a representative example we choose the iconic one-dimensional hard
rod system, for which analytical solutions exist \cite{percus1976,
  robledo1981}. To induce spatial inhomogeneity we consider
confinement between two hard walls. The hard core nature of this test
situation poses a stringent test for the gauge invariance as the
finite particle shifting will in general lead to overlapping particle
configurations and consequentially to a differing sequence of
microstates in the Monte Carlo Markov chain.  The one-dimensional
shifting field is chosen as either $\epsilon(x_i)=0$, which reproduces
the original system and constitutes our reference, or
$\epsilon(x_i)=0.5\sin(4\pi x_i/L)$, where $L=10a$ is the separation
distance between the two hard walls, $a$ is the particle diameter, and
$x_i$ is the one-dimensional position coordinate of particle
$i$. According to Eqs.~\eqref{LQriTrafo} and \eqref{LQpviTrafo} the
transformed variables are $\tilde x_i = x_i + \epsilon(x_i)$ and
$\tilde p_i= [1+\partial \epsilon(x_i)/\partial x_i]^{-1}p_i$. The
construction of Monte Carlo trial moves is identical in the shifted
and unshifted sytems, in that $x_i$ and $p_i$ are displaced uniformly
within a maximal cutoff. In the unshifted system the new trial state
enters the Boltzmann factor to accept or reject the new state
according to the Metropolis criterion \cite{frenkel2023book}, as is
the standard procedure. In the shifted system, the Boltzmann factor is
evaluated on the basis of the shifted variables $\tilde x_i$ and
$\tilde p_i$, both before and after the trial move. As a
representative observable, we show in Fig.~\ref{FIGshiftingMC} results
for the density profile $\rho(x)$ as obtained from histograms of
particle positions $\tilde x_i$ from separate runs without and with
shifting. We also show the position- and momentum-resolved one-body
phase space density $f(x,p)$, where $p$ denotes the momentum variable.
Consistent with the theoretical structure of the particle gauge
invariance, the results in the shifted system are identical to those
in the original system with $p$ following the correct Maxwellian.  We
have ascertained that the same behaviour holds when replacing the hard
core wall and interparticle potentials by soft potentials, using the
Lennard-Jones form as representative.  Results are shown in
Appendix~\ref{SECdoubleWell} together with a confirmation of the gauge
invariance in a double well.  As expected the invariance holds already
canonically with fixed $N$.

We have restricted ourselves to forms of phase space shifting given by
Eqs.~\eqref{LQriTrafo} and \eqref{LQpviTrafo} together with
Hamiltonians \eqref{LQHamiltonian} that feature standard kinetic
energy. The position transform~\eqref{LQriTrafo} is a general and
freely chosen diffeomorphism. The specific form of the momentum
transform~\eqref{LQpviTrafo} then follows uniquely from imposing i)
that the transformation is canonical (and hence the Jacobian is unity)
and ii) that the identity transformation is recovered for
$\eps(\rv)=0$. We leave investigations of possible generalizations to
future work. The relationship of the present theory with existing sum
rules is recapped in Appendix~\ref{SECsumRules}.  As the shifting
operators \eqref{LQsigma} feature a sum over all particles, with
identical terms that involve only a single particle $i$, the
statistical mechanical symmetry with respect to particle index
permutation is respected. Hence applying $\bsig(\rv)$ commutes with
index permutation.

In conclusion we have shown that statistical mechanical microstates
carry an intrinsic ambiguity with respect to the gauge shifting
transformation \eqref{LQriTrafo} and \eqref{LQpviTrafo}. The Lie
algebra \eqref{LQsigmaAlgebra} for infinitesimal generators
\eqref{LQsigma} is imprinted in measurable physical correlation
functions.  Numerical implementation of the finite shifting gives
additional freedom for particle-based simulation techniques and one
can envisage rich cross fertilization with force sampling schemes
\cite{borgis2013, delasheras2018forceSampling, coles2019, coles2021,
  rotenberg2020, purohit2019} and the mapped averaging
framework~\cite{schultz2016, trokhymchuk2019, schultz2019,
  purohit2019}. Recent progress in microscopy-based measurement of
locally resolved forces in colloidal systems \cite{dong2022} offers
exciting potential for carrying out corresponding experimental work.

\acknowledgments We thank Thomas Kriecherbauer and Vollrath Martin Axt
for useful discussions.

\section*{End Matter}

\subsection{Relationship to functional methods}
\label{SECrelationshipToFunctionalMethods}

We can identify the effect of applying $\bsig(\rv)$ to a generic phase
space function $f(\rv^N,\pv^N)$ to be equivalent to the following
functional derivatives of the considered function in the transformed
variables:
\begin{align}
  \bsig(\rv) f(\rv^N,\pv^N)
  &=  \frac{\delta f(\tilde\rv^N, \tilde\pv^N)}{\delta\eps(\rv)} 
  \Big|_{\eps=0},
  \label{LQsigmaAsFunctionalDerivative}\\
  \bsig(\rv) \bsig(\rv') f(\rv^N,\pv^N) &= 
  \frac{\delta^2 f(\tilde\rv^N, \tilde\pv^N)}
       {\delta\eps(\rv)\delta\eps(\rv')} \Big|_{\eps=0}
  \notag\\&\quad
  +[\nabla\delta(\rv-\rv')] \bsig(\rv) f(\rv^N, \pv^N).
  \label{LQdoublesigmaAsFunctionalDerivative}
\end{align}
We recall that the tilde indicates the transformed phase space
variables~\eqref{LQriTrafo} and \eqref{LQpviTrafo}.  Equation
\eqref{LQsigmaAsFunctionalDerivative} follows from differentiating
\eqr{LQfShiftedBySigma} and the analogous second-order version
\eqr{LQdoublesigmaAsFunctionalDerivative} follows iteratively.

\subsection{Gauge invariance for Lennard-Jones particles}
\label{SECdoubleWell}

\begin{figure*}[!t]
  \vspace{1mm}
 \includegraphics[width=.32\textwidth]{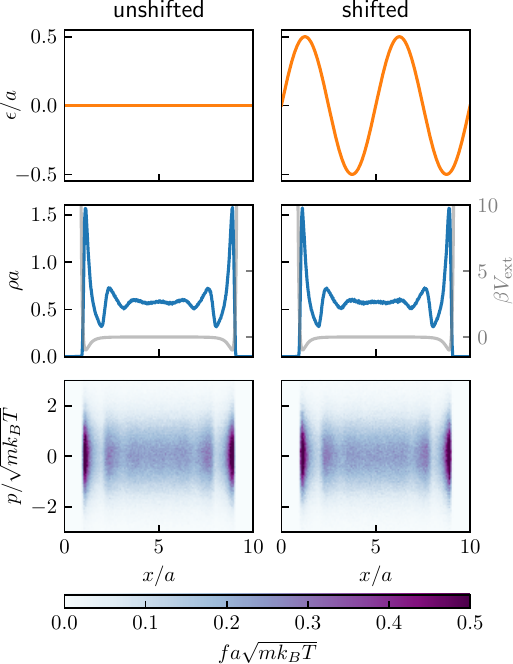}
 \includegraphics[width=.32\textwidth]{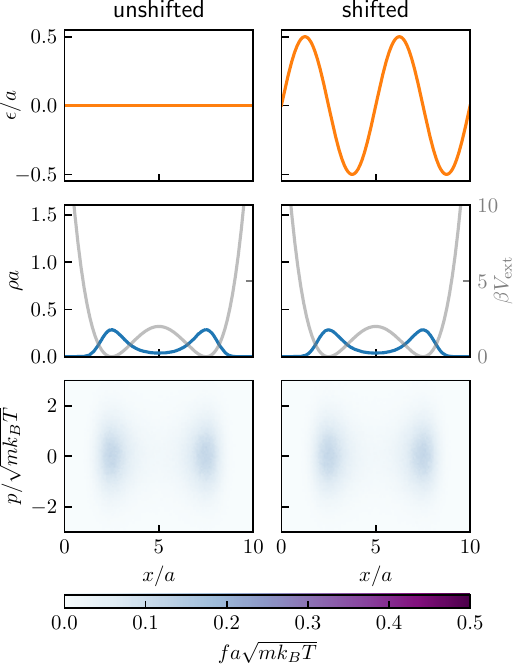}
 \includegraphics[width=.32\textwidth]{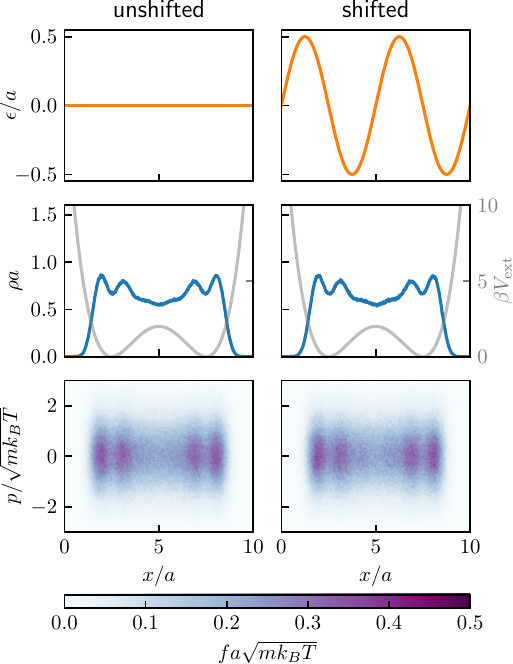}
 \caption{Computer simulation results for the gauge invariance of
   Lennard-Jones particles. Shown are three cases: i) confinement
   between two Lennard-Jones walls (first and second column), ii)
   trapping inside of a double well external potential for particle
   number $N=1$ (third and fourth column), and iii) trapping in the
   double well for $N=5$ particles (fifth and sixth column). Monte
   Carlo results for the original unshifted system are shown as a
   reference (first, third, and fifth columns). The simulation results
   for the scaled density profile $\rho(x)a$ (middle panels) and for
   the one-body phase space distribution function
   $f(x,p)a\sqrt{mk_BT}$ (bottom panels) in the shifted system
   (second, fourth, and sixth columns) are numerically identical to
   the respective results in the unshifted system.}
\label{FIGdoubleWell}
\end{figure*}

To explicitly demonstrate that the gauge invariance holds beyond the
hard core case, shown in Fig.~\ref{FIGshiftingMC}, we consider
confinement of Lennard-Jones particles between two Lennard-Jones
walls, as combined from the single wall potential $\beta V_\rmext(x) =
4[(a/r)^{12} - (a/r)^6]$.  We also consider trapping in a double well
with barrier height $e_w$ and separation $2x_m$ between the two
potential minima. In this case the external potential is $V_\rmext(x)
= e_w [(x - L/2)^2 - x_m^2]^2 / x_m^4$, where $L$ is the system
length, the barrier is located at $L/2$, and we set $\beta e_w=2$ and
$x_m = 2.5a$.
Monte Carlo results for the behaviour both of a single particle
$(N=1)$ and for $N=5$ particles are shown in
Fig.~\ref{FIGdoubleWell}. The shifting field $\epsilon(x)$ has the
sinusoidal form described in the main text. We find that the results
for the shifted and the original system are numerically identical, see
Fig.~\ref{FIGdoubleWell}.

\subsection{Relationship to exisiting sum rules}
\label{SECsumRules}

Several of the sum rules that arise from the present statistical
mechanical gauge invariance possess close relationships to the liquid
state literature \cite{hansen2013, evans1979, baus1984, evans1990,
  henderson1992, triezenberg1972}. For sum rules that follow from
global shifting invariance we refer the reader to the description in
the Methods section of Ref.~\cite{hermann2021noether}.  Noether
two-body force correlation identities such as the ``3g''-sum rule
\cite{sammueller2023whatIsLiquid, hermann2023whatIsLiquid} can
alternatively to invariance arguments be obtained from partial
integration on phase space, as described in the Appendix of
Ref.~\cite{hermann2023whatIsLiquid}. Analogously, global hyperforce
sum rules can be derived from Yvon's or from Hirschfelder's theorem,
as described in Ref.~\cite{robitschko2024any}. Investigating the
relationship of the gauge invariance to the (nonequilibrium)
fluctuation theorems of stochastic thermodynamics \cite{seifert2012}
constitutes a highly valuable goal for future work.


\end{document}